\documentclass[12pt]{article}
\usepackage{times}
\usepackage{geometry}
\usepackage[utf8]{inputenc}
\usepackage{enumitem,amssymb}
\usepackage{ragged2e}
\usepackage{graphicx}
\usepackage{wrapfig}
\usepackage{natbib}
\newcommand{\aap}{A\&A}

\newcommand{\apjs}{ApJSS}

\newcommand{\apjl}{ApJ}

\newlist{thematic}{itemize}{8}
\setlist[thematic]{label=$\square$}
\usepackage{pifont}
\newcommand{\cmark}{\ding{51}}%
\newcommand{\done}{\rlap{$\square$}{\raisebox{2pt}{\large\hspace{1pt}\cmark}}%
\hspace{-2.5pt}}

\begin{document}
\raggedright
\huge
Astro2020 Science White Paper \linebreak

The Origin of Elements Across Cosmic Time \linebreak
\normalsize

\noindent \textbf{Thematic Areas:} \hspace*{60pt} $\square$ Planetary Systems \hspace*{10pt} $\square$ Star and Planet Formation \hspace*{20pt}\linebreak
$\done$ Formation and Evolution of Compact Objects \hspace*{31pt} $\square$ Cosmology and Fundamental Physics \linebreak
  $\done$  Stars and Stellar Evolution \hspace*{1pt} $\square$ Resolved Stellar Populations and their Environments \hspace*{40pt} \linebreak
  $\square$    Galaxy Evolution   \hspace*{45pt} $\square$             Multi-Messenger Astronomy and Astrophysics \hspace*{65pt} \linebreak
  
\textbf{Principal Author:}

Name: Jennifer Johnson	
 \linebreak						
Institution: Department of Astronomy, Ohio State University
 \linebreak
Email: johnson.3064@osu.edu
 \linebreak
Phone:  614-292-5651
 \linebreak
 
\textbf{Co-authors:} {\bf Gail Zasowski} (University of Utah), {\bf David Weinberg} (Ohio State University), {\bf Yuan-Sen Ting} (Institute for Advanced Study/Princeton University/OCIW), {\bf Jennifer Sobeck} (University of Washington), {\bf Verne Smith} (NOAO), {\bf Victor Silva Aguirre} (Aarhus University),  {\bf David Nataf} (Johns Hopkins), {\bf Sara Lucatello} (INAF Osservatorio Astronomico di Padova), {\bf Juna Kollmeier} (OCIW), {\bf Saskia Hekker} (Max Planck Institute for Solar System Research), {\bf Katia Cunha} (University of Arizona), {\bf Cristina Chiappini} (AIP),  {\bf Joleen Carlberg} (STScI), {\bf Jonathan Bird} (Vanderbilt University), {\bf Sarbani Basu} (Yale University), {\bf Borja Anguiano} (University of Virginia)
  \linebreak

\textbf{Abstract:}

The problem of the origin of the elements is a fundamental one in astronomy and one that has many open questions. Prominent examples include (1) the nature of Type Ia supernovae and the timescale of their contributions; (2) the observational identification of elements such as titanium and potassium with the $\alpha$-elements in conflict with core-collapse supernova predictions; (3) the number and relative importance of r-process sites; (4) the origin of carbon and nitrogen and the influence of mixing and mass loss in winds; and (5) the origin of the intermediate elements, such as
Cu, Ge, As, and Se, that bridge the region between charged-particle and neutron-capture reactions. The next decade will bring to maturity many of the new tools that have recently made their mark, such as large-scale chemical cartography of the Milky Way and its satellites, the addition of astrometric and asteroseismic information, the detection and characterization of gravitational wave events, 3-D simulations of convection and model atmospheres, and improved laboratory measurements for transition probabilities and nuclear masses. All of these areas are key for continued improvement, and such improvement will benefit many areas of astrophysics. 

\pagebreak

\section{Introduction}

The origin of the elements remains one of the great problems in astrophysics
and nuclear physics.  We understand some basic components of the story,
such as the origin of $\alpha$-elements in core collapse supernovae (CCSN)
and the contribution of both CCSN and Type Ia supernovae (SNIa) to
iron-peak elements. However, our understanding of many other elements
is more uncertain, and we have only limited knowledge of where and when
the elements in present-day Milky Way stars were formed and of how many
metals of different species were ejected from the Galaxy rather than
incorporated into stars. The elemental abundances of stars and the Interstellar Medium (ISM),
in the Milky Way and other galaxies, are sensitively tied to many other
areas of astrophysics: the physics of all varieties of supernovae;
rotation and internal mixing in massive stars and intermediate mass stars;
binary stellar evolution; the stellar initial mass function and its
variation with metallicity, time, or environment; black hole formation;
neutron star mergers; the assembly and star formation history of the
Milky Way; the physics of galaxy formation, star formation, and feedback;
the formation and composition of planets.

The study of galactic chemical evolution has made dramatic strides over
the past decade thanks largely to massive spectroscopic surveys that
measure abundances and in some cases detailed chemical fingerprints of $\sim 10^5$ stars. These surveys have demanded radical
advances in analysis techniques, and they have been complemented by
asteroseismology from {\it Kepler}, parallaxes and proper motions
from {\it Gaia}, IFU abundance maps of thousands of low redshift galaxies,
measurements of gas content and metallicity evolution out to $z=4$,
and high-resolution spectroscopy of growing samples of selected stellar
populations (low metallicity, solar twins, dwarf satellites and streams)
from large telescopes. The 2020s should see big advances on all of these
fronts and an influx of critical new information from time-domain surveys
(probing supernova progenitors and black hole formation) and from
gravitational wave discoveries.

\section{Key Science and Open Questions}

{\bf Oxygen through the iron peak}
One of the challenges in determining the origin of elements is separating
the impact of nucleosynthesis physics from the accretion, star formation,
and outflow history of the Galaxy.  A recent study from the SDSS APOGEE
survey \citep{weinberg2019} shows that one can simplify interpretation
by taking Mg as a reference element and separately analyzing high-$\alpha$
and low-$\alpha$ stellar populations, which have different relative SNIa contributions for all elements.  The median sequences of  
[X/Mg] vs. [Mg/H] are independent of location in 
the Galactic disk for X = O, Si, S, Ca, Na, Al, P, K, V, Cr, Mn, Fe,
Co, Ni, indicating that these sequences trace IMF-averaged yields
with little sensitivity to details of evolution.

Fig.~1 compares the empirically inferred values of $f_{\rm cc}$ --- the
fraction of each element contributed by CCSN at [X/Mg] = [Mg/H] = 0 ---
to predictions for the two different CCSN+SNIa+AGB compilations  
incorporated in the {\it Chempy} code of \citet{rybizki2017}.
For some elements there is good agreement of observation and theory,
for others (e.g., Na, P) there is dramatic discrepancy, and near the
iron peak there are large disagreements between these two yield compilations.
In the near future this approach can be extended to include new elements
from APOGEE and optical surveys such as GALAH, to new stellar populations
in the bulge, halo, and dwarf satellites, and to investigate star-by-star
deviations from median trends.  These results will provide strong tests
of supernova and AGB nucleosynthesis models and probes of IMF variations.

\begin{figure}[th]
\includegraphics[width=\textwidth]{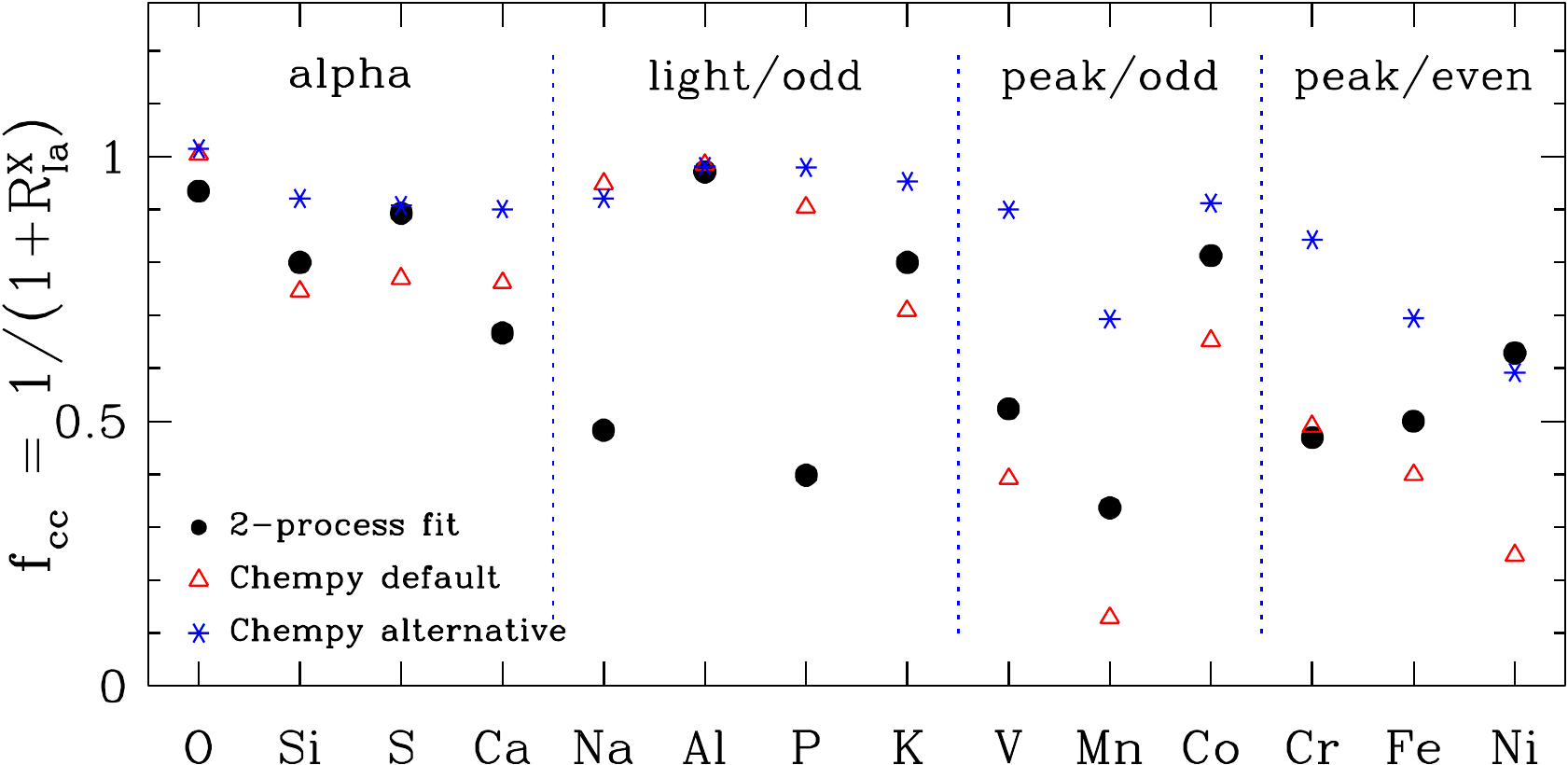}
\caption{
Fractional contribution of CCSN to the abundance of elements
(marked on the $x$-axis) at solar abundances [X/Mg] = [Mg/H] = 0.
Filled circles show observational inferences from APOGEE, while
triangles and asterisks show the predictions for a single stellar 
population of solar metallicity from the default and alternative
yield compilations of \citet{rybizki2017}.  Adapted from \citet{weinberg2019}.
}
\end{figure}

{\bf Carbon and Nitrogen}
Carbon and nitrogen are ejected by massive star winds, by CCSN and by AGB winds. The relative contribution of these three mechanisms, as a function of metallicity and in
different environments, is poorly understood.  The approach illustrated
in Fig.~1 is one promising way forward. Surface C and N abundances
in red giants reflect internal evolution as well as birth abundances,
which makes C/N a valuable spectroscopic age diagnostic \citep[e.g.,][]{martig2016} but
complicates interpretation of red giant surveys. Reliable determinations
of CCSN vs. AGB yields would provide crucial constraints on mixing
processes in supernova progenitors and red giants.  They are also
essential for interpreting gas phase ISM abundances, for which
C, N, and O are among the best measured elements and whose abundances cannot
be explained by simple models of nitrogen as a pure secondary element \citep[e.g.,][]{vangioni_2018}.

\vspace{0.07in}
{\bf Contributions of neutron star mergers in the production of the r-process}
It has long been recognized that typical CCSN do not have the physical
conditions needed to produce the heavier $r$-process elements such as
europium \citep{thompson2001}.  Neutron star mergers have been proposed as
an alternative site \citep{lattimer1974}, and this idea has gained
currency first with evidence that $r$-process enrichment in low mass
galaxies is highly stochastic \citep[e.g.,][]{ji_reticulum} and more dramatically with
spectroscopic demonstration of heavy element production in the neutron
star merger GW170817 \citep{kilpatrick_2017}.  But we do not know whether neutron star mergers dominate the production of all $r$-process elements in all environments,
or whether typical CCSN or rare categories of massive star explosions
\citep[e.g.,][]{thompson2003,siegel2018} make important contributions, at least for the 1st-peak and 2nd-peak elements. 
There is evidence that the production of the light neutron-capture elements is independent of the heavy elements
\begin{wrapfigure}{r}{0.4\textwidth}
\begin{center}
\includegraphics[width=0.38\textwidth]{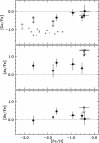}
\end{center}
\caption{Figure from \citet{roederer2014} showing the [Se, As, Ge/Fe] ratios in metal-poor stars. These Z=32-34 elements show surprising deviations from solar ratios; the origin of these ratios is not yet understood.}
\end{wrapfigure}
\citep[e.g.,][]{travaglio2004}, leading to the idea of a "weak r-process" \citep[e.g.,][]{honda2006}. We expect progress in
the coming decade from more systematic multi-element studies of low
metallicity stars in the halo and dwarf satellites and from spectroscopic
studies of more neutron star mergers discovered as gravitational wave
transients.  The results will provide strong tests of theoretical studies
of neutron star mergers and 3-dimensional massive star explosions.

\vspace{0.07in}
{\bf Production of the Z=29-50 elements}

The origin of the elements from copper to first r-process peak at Z$\sim$ 50 is the most poorly understood area of nucleosynthesis (see Figure 2). They are both the heaviest elements that can be produced by charged particle reactions and the lightest elements where substantial contributions from neutron captures are expected. Substantial contributions are also expected from the weak s-process in helium burning in massive stars and from the main s-process in AGB stars. The presence of large enhancements in some of these elements in actinide-rich stars \citep{sneden2003} and the failure of r- and s-process models to match the observed patterns in metal-poor stars \citep{roederer_i} provide intriguing clues that have not yet been deciphered. Disentangling these process to glimpse how far charged particle reactions can go in CCSNe and when neutron captures become dominant is a task for the 2020s.

\section{Impact Across Astrophysics}

{\bf Proxies for fundamental properties} 
Both the overall enrichment and the ratios of key elements are correlated with the age of a star \citep{tinsley1979}. Turning a relative ranking of stellar ages into an absolute age since star formation began is not yet achievable. However, the measurement of accurate absolute ages for a large number of field stars using Gaia parallaxes \citep[e.g.,][]{delgado2019} or asteroseismology \citep[e.g.,][]{silva2018} has made the robust calibration of chronometers feasible over the next decade. One of the most important results from this work will be the firm establishment of the timescale for SNIa to contribute substantial amounts of iron, a number that is currently based on a specific chemical evolution model \citep{matteucci1986}.
In addition, the promising method to measure ages for red giant stars throughout the Galaxy using their post first-dredge up C/N ratios \citep{martig2016} requires accurate chemical evolution of C and N to have the correct initial pre-dredge up values \citep{shetrone2019}.\\

\vspace{0.07in}
{\bf Physics of stars}
Because nucleosynthesis is sensitive to conditions in stellar interiors, matching the observed abundances requires a nuanced understanding of many physical phenomena. For example, nitrogen production in massive stars cannot be correctly predicted without modeling rotationally induced mixing and the extent of winds in the pre-supernova phase; likewise nitrogen production in hot-bottom burning in AGB stars means understanding convection and instabilities during third dredge-up. The presence of a close companion, a common occurrence for massive stars in particular \citep[e.g.,][]{moe2017} also affects stellar evolution and will require suitable modeling. The combination of observed abundances, deeper understanding of stellar interiors from seismology, and theoretical modeling will open up new pathways to understand the most complicated parts of stellar evolution. These include the production of abundance anomalies in globular clusters \citep[e.g.,][]{ggc} and the depletion of Li in hot metal-poor stars below the value predicted from Big Bang nucleosynthesis \citep[e.g.,][]{li}.
\\

\vspace{0.07in}
{\bf Fundamental properties of heavy nuclei} 
Adequate theories of nuclear structure do not exist for heavy nuclei near the neutron-drip line. Even masses, the most basic property, are not known; reliable theoretical or laboratory information on more complex information such as neutron-capture cross-sections is even more distant. Astrophysical measurements of abundances produced in the r-process cannot distinguish between patterns imposed by nuclear physics and patterns imposed by the physical conditions where the neutron capture occur. If the theoretical conditions of the r-process site can be reliably established, meaningful constraints can be put on the properties of the most neutron-rich nuclei \citep[e.g.,][]{mumpower2016}. 

\vspace{0.07in}
{\bf Appearance of planet-building material} 
The frequency of formation of planets, at least for certain masses, depends on the composition of the protoplanetary nebula \citep{Valenti08}. The abundance ratios of the refractory elements can have profound impact on structure of rocky planets \citep[e.g.][]{Unterborn17} and the amount of long-lived radioactive elements, in particular K, Th, and U, provides energy to keep planetary interiors hot and increases the likelihood of plate tectonics and the accompanying climate regulation \citep{Unterborn14}. 

\vspace{0.07in}
{\bf Nature of Population III Stars} 
No stars composed purely of Big Bang material have yet been found. If current theories that no long-lived low-mass stars can be formed from gas devoid of carbon and oxygen are correct \citep[][but see \citet{schlaufman2018}]{frebel2007}, no such star will ever be found. The explosions of Population III stars may not be easily visible, even with the next generation of space telescopes \citep{tumlinson}. This leaves their enrichment of the gas that formed the next generation of stars as the most promising observational probe of their nature. Stars with [Fe/H] $ < -3$ frequently show enhanced carbon abundances and other abundance anomalies compared to their slightly more enriched kin, hinting at nucleosynthetic phenomena confined to the early Universe \citep{frebel_norris}.

\section{Capabilities in the 2020s and Beyond}
Remarkable progress was originally made by considering solar abundances, the chemical evolution of the solar neighborhood, and special stars showing large overabundances from a specific process.
Measurements of multi-element abundances across the Galaxy and into neighboring galaxies are opening up new environments with distinct enrichment histories and revealing examples of rare stars that highlight specific processes.\\

\subsection{Observational Needs}

{\bf Industrial scale Spectroscopy of the Milky Way and its Satellites}
To identify crucial test cases where one nucleosynthetic site has a dominant presence and to understand the overall chemical evolution of the Galaxy and its environs requires observing millions of stars at medium resolution. To measure the abundances of many elements in stars in different environments across a wide range of ages will require high-resolution, high-SNR spectra from the ultraviolet, where strong lines of heavy elements are present, to the infrared, where the best measurements for C and O can be made from molecular lines. \\

\vspace{0.07in}
{\bf Gravitational Wave Astronomy}
The revolution in astronomy in the next decades from the routine observation of 
gravitational radiation will reveal the frequency of neutron star mergers as well as the number and expected merging times of close white dwarf binaries. With this information, the amount of heavy $r$-process material produced per neutron star merger and the dominance of the double-degenerate model for SNIa can be established.

\vspace{0.07in}
{\bf Supernovae and their progenitors}
The results of transient surveys, such as ASAS-SN \citep{Shappee14} and LSST \citep{Tyson03}, will provide counts of CCSNe and SNIa in the relatively nearby Universe; spectroscopic follow-up is required for accurate typing.  An accurate normalization and shape of SNIa delay time distribution is an essential ingredient for chemical evolution models.

\vspace{0.07in}
{\bf Asteroseismology} Asteroseismology probes the conditions in stellar interiors and produces ages and evolutionary states. The high-precision space-based photometry missions of the past decade have started the revolution. A continued supply of asteroseismic data for stars across the Galaxy is a high priority in the next decade.

\subsection{Theoretical and Modeling Needs}
A robust modeling effort in the models of pre-supernova stars and  mechanisms of CCSNe and SNIa explosions and the effect on the resulting nucleosynthesis is critical in the upcoming decade. Attempts to reconcile the discrepancies shown in Figure 1 for the iron peak elements will probe SNIa yields, and therefore single- vs. double-degenerate models. For lower mass stars, accurate modeling of convection and mixing through the AGB phase has begun on small scales \citep{Herwig13}; these will be needed to be expanded and additional physical effects included. Deriving accurate abundances from stellar spectra requires continued development of realistic 3-D model atmospheres.

\subsection{Laboratory Astrophysics}
The study of the origin of the elements depends on laboratory astrophysics, including accurate atomic parameters  appropriate for a wide range of transition probabilities and wavelengths. Among the critical needs are near-infrared transitions \citep{Shetrone15}, accurate compilations of molecular data \citep[e.g.,][]{Souto17}, detailed cross-sections for reliable non-LTE calculations \citep[see e.g.][]{Alexeeva18}. Measurement of cross-sections for nuclear reactions and of masses for neutron-rich nuclei are important as well. Continued funding for laboratory astrophysics remains a high-priority for advances in the origin of the elements.

\pagebreak

\bibliography{wp}

\end{document}